\documentclass[aps,prc,twocolumn,floatfix,superscriptaddress,nofootinbib]{revtex4}

\usepackage{epsfig}
\usepackage{amsmath}
\usepackage{color}

\usepackage{graphicx}

\begin{document}

\title{Collision centrality and $\tau_0$ dependence of the emission of thermal photons from fluctuating initial state in ideal hydrodynamic calculation}
\author{Rupa Chatterjee}
\email{rupa.r.chatterjee@jyu.fi}
\affiliation{Department of Physics, P.O.Box 35, FI-40014 University of Jyv\"askyl\"a, Finland}
\author{Hannu Holopainen}
\email{holopainen@fias.uni-frankfurt.de}
\affiliation{Department of Physics, P.O.Box 35, FI-40014 University of Jyv\"askyl\"a, Finland}
\affiliation{Helsinki Institute of Physics, P.O.Box 64, FI-00014 University of Helsinki, Finland}
\affiliation{Frankfurt Institute for Advanced Studies, Ruth-Moufang-Str. 1, D-60438 Frankfurt am Main, Germany}
\author{Thorsten Renk}
\email{thorsten.renk@pjys.jyu.fi}
\affiliation{Department of Physics, P.O.Box 35, FI-40014 University of Jyv\"askyl\"a, Finland}
\affiliation{Helsinki Institute of Physics, P.O.Box 64, FI-00014 University of Helsinki, Finland}
\author{Kari J. Eskola}
\email{kari.eskola@phys.jyu.fi}
\affiliation{Department of Physics, P.O.Box 35, FI-40014 University of Jyv\"askyl\"a, Finland}
\affiliation{Helsinki Institute of Physics, P.O.Box 64, FI-00014 University of Helsinki, Finland}

\begin{abstract}
Fluctuations in the initial QCD matter density distribution  are found to enhance the production of thermal photons significantly in the range $2 \le p_T \le 4$ GeV/$c$  compared to a smooth initial state averaged profile in ideal hydrodynamic calculation for 200 AGeV Au+Au collisions at the Relativistic Heavy Ion Collider (RHIC) and 2.76 ATeV Pb+Pb collisions at the Large Hadron Collider (LHC).  The thermal emission of photons is strongly dependent on the initial temperature of the system where the presence of   
 'hotspots' in the initial state translates into enhanced production of photons compared to a smooth profile. The effect of fluctuations in the initial state  is found to be stronger for peripheral collisions and for lower beam energies. The $p_T$ spectra are found to be quite sensitive to the value of the initial formation time of the plasma which is not known unambiguously and which may vary with collision centralities at a particular beam energy. 
Increase in the value of the formation time lowers the production of thermal photons compared to the results from a shorter  formation time. 
However, the relative enhancement from fluctuating initial states (compared to a smooth initial state) is found to be stronger for the larger values of formation time. The $p_T$  spectra alone are found  to be  insufficient to quantify the fluctuations in the initial density distribution due to the uncertainties in the initial conditions.  A suitably normalized ratio of central-to-peripheral yield as a function of collision centrality and $p_T$ can be a useful measure  of the fluctuation size scale.

\end{abstract}

\maketitle
\section{Introduction}
Ideal hydrodynamics with smooth initial density distribution has been used successfully in recent past to model the evolution of the hot and dense system produced in collisions of heavy nuclei at relativistic energies. The experimental data~\cite{fl2} for $p_T$ spectra and elliptic flow of several hadrons at the Relativistic Heavy Ion Collider (RHIC) energy are well reproduced upto a relatively large value of transverse momentum ($p_T \sim 2$ GeV/$c$) using hydrodynamic models~\cite{fl1,hydro,Niemi:2008ta, hirano, uli} with smooth initial conditions (IC). However, for any given single event there may be randomly distributed inhomogenities in the initial energy density. 
 Lately it has been shown that event-by-event (e-by-e) fluctuations in the IC have sizable effect on the physical observables if they are computed by averaging over many final states rather than considering an averaged initial density profile.

Recent studies have reported that e-by-e  fluctuating IC reproduces the experimental charged particle elliptic flow even for the most central collisions at RHIC~\cite{hannu} which was underestimated by all earlier hydrodynamic calculations using smooth IC. The e-by-e hydrodynamics with fluctuations in the IC also give a better agreement of the experimental charged particle spectra at high $p_T$ by increasing the number of particles there~\cite{hama, hannu}. In addition, the initial state fluctuations may also help to understand the various structures observed in two-particle correlations~\cite{andrade}.

Direct photons are considered as one of the few observables able to probe the initial state of the hot and dense matter produced in the collisions.  Thermal emission of photons from the evolving system shows a very strong temperature dependence. 
Photons having $p_T \ > \ 1 \ \rm{GeV/}$$c$ are expected to provide a glimpse of the early part of the expansion history. These photons are especially suitable for probing the fluctuations in the IC due to their sensitivity to the initial temperature of the plasma~\cite{phot}.

Thermal radiation is predicted to be the dominant source of direct photons in the range $1 \le p_T \le 3$ GeV/$c$~\cite{phenix1,dks_qm08}. However, photon results using smooth IC and latest rates~\cite{amy, trg} fall well below the experimental data points (Fig. 4 of Ref.~\cite{phenix1}) in this $p_T$ range. In a recent work~\cite{chre} we have shown a substantial enhancement in the production of thermal photons for $p_T  > 1$ GeV/$c$ with an e-by-e fluctuating IC relative to a smooth initial-state averaged profile in the ideal hydrodynamic calculation for 200 AGeV Au+Au collisions at RHIC and for the 0--20\% centrality bin. Consequently, our results with fluctuating density profile improve the agreement with the PHENIX experimental data for $2 < p_T < 4$ GeV/$c$ in that centrality bin. The enhancement was found to scale with the inverse of the initial fluctuation size.

Here we perform a more systematic study of the effects of the initial state fluctuations on the production of thermal photons in heavy ion collisions and propose an experimentally measurable quantity $R_{cp}^\gamma$ that can be useful to quantify the fluctuation size realized in nature.

The effects of the initial density fluctuations are found to be more pronounced for lower beam energies and for smaller size systems. In addition, the initial formation time of the plasma plays a crucial role in all these calculations. As the value of $\tau_0$ is not known unambiguously, we see that it is difficult to distinguish between the results from fluctuating IC at a particular $\tau_0$ and the results at a relatively smaller $\tau_0$  using the smooth IC. This is relevant since one can argue that the formation time of the plasma for peripheral collisions may not remain the same as for central collisions since the initial energy density and system size become smaller moving from central to peripheral collisions. 

The effects of  the initial state fluctuations on the production of thermal photons are studied with different values of $\tau_0$ changing with collision centrality at the LHC energy. The centrality dependence of $\tau_0$ is taken from the EKRT model calculations~\cite{Eskola:1999fc, risto}. Larger values of $\tau_0$ for peripheral collisions increase the difference between the results from the smooth and the fluctuating IC slightly compared to the results with a smaller $\tau_0$. From  the centrality dependent results at different formation times we conclude that we need to look for a quantity besides the photon spectra as these  are insufficient to quantify the fluctuations in the IC.  
  The ratio of central to peripheral yield normalized by the number of binary collisions, $R_{cp}^{\gamma}$, shows potential to probe the density fluctuations and their size in the IC.

The paper is organized as follows. First we briefly describe the e-by-e  hydrodynamic framework used in this study and the production of thermal photons from it. In the next section photon results are shown from the smooth and fluctuating IC at RHIC for different collision centralities. 
 After that  we show the thermal photon $p_T$ spectra from  the smooth and fluctuating IC at the LHC energy and compare those with the RHIC results.
 Photon results from the fluctuating IC considering different values of the initial formation time $\tau_0$ of the plasma at RHIC are shown in the next section along with the results from centrality dependent $\tau_0$ values at LHC. 
 Next we define  the ratio $R_{cp}^{\gamma}$ for thermal photons and calculate it   as a function of collision centrality and the fluctuation size for different values of transverse momentum.     
Finally we summarize our results.

\section{Thermal photons from event-by-event hydrodynamics}

\subsection{Event-by-event hydrodynamics}

We utilize here the e-by-e hydrodynamical model developed in~\cite{hannu}, which has been
applied successfully to study the hadronic spectra and elliptic flow with fluctuating IC,
 and to calculate the thermal photon emission at RHIC~\cite{chre} and LHC energies. The model is based on
ideal hydrodynamics with the assumption of longitudinal boost invariance and the remaining
(2+1)-dimensional problem is solved numerically using the SHASTA algorithm
\cite{Boris, Zalesak}. The Equation of State (EoS), which shows a sharp crossover
transition from the plasma phase to hadronic matter, is taken from~\cite{Laine:2006cp}. 

The IC are obtained from the Monte Carlo Glauber (MCG) model. The standard two-parameter
Woods-Saxon nuclear density profile is used to randomly distribute the nucleons in the
colliding nuclei. Thus no nucleon-nucleon correlations or finite nucleon size effects are
taken into account. Nucleons $i$ and $j$ from two different nuclei are assumed to collide
whenever they satisfy the relation,
\begin{equation}
  (x_i - x_j)^2 + (y_i-y_j)^2 \le \frac{\sigma_{NN}}{\pi},
\end{equation}
where $\sigma_{NN}$ is the inelastic nucleon-nucleon cross section and $x_i,y_i$ denote
the positions of the nucleons in the transverse plane. The centrality classes are defined by binning the
distribution of events in the number of participants
($N_{\rm {part}}$) as in Ref.~\cite{hannu}. Table~\ref{table1.1} shows the $N_ {\rm {part}}$ ranges
corresponding to the different centrality classes, the average value of impact parameter
and average number of participants for each centrality class for 200  AGeV Au+Au
collisions at RHIC. Table~\ref{table1.2} shows the same for 2.76  ATeV Pb+Pb
collisions at LHC. The inelastic nucleon-nucleon cross section is taken as 42 (64) mb for $\sqrt{s_{NN}}=$ 0.2 (2.76)  TeV.

\begin{table}
\begin{tabular}{ | c | c | c | c |c |}
 \hline
 Centrality \% & $N_{\rm {part}}$ range & $\langle b \rangle  $ (fm) &\hspace{.2 cm} $\langle N_{\rm {part}} \rangle$ &  $\langle N_{\rm {bin}} \rangle$\\ \hline
	\hline
      0 -- 20 & 394 -- 197 & 4.44 & 280.0 & 778.3\\ \hline
     20 -- 40 & 196 -- 93 &  8.08 & 139.7 & 293.6 \\  \hline
     40 -- 60 & 92 -- 35  & 10.5 &  60.24 & 90.48 \\ \hline
 \end{tabular}
\caption{\label{table1.1}  The $N_{\rm {part}}$ range, average impact parameter  $\langle b \rangle $, average numbers of participants and average numbers of binary collisions for different centrality classes of  200  AGeV Au+Au collisions at RHIC according to the MC Glauber model with $\sigma_{NN}=$ 42 mb.}

\begin{center}
    \begin{tabular}{ | c | c | c | c | c |}
    \hline
     Centrality \% & $N_{\rm {part}}$ range & $\langle b \rangle  $ (fm) & $\langle N_{\rm {part}} \rangle$ & $\langle N_{\rm {bin}} \rangle$ \\ \hline
     \hline
      0 -- 20 & 416 -- 221 & 4.64 & 309.3 & 1217.0\\ \hline
     20 -- 40 & 220 -- 104 & 8.52 & 157.0 &  440.9 \\\hline
     40 -- 60 & 103 -- 39  & 11.1 & 67.20 &  125.2 \\\hline
    \end{tabular}	
\caption{The $N_{\rm {part}}$ range, average impact parameter  $\langle b \rangle $, average numbers of participants and average numbers of binary collisions for different centrality classes of 2.76  ATeV Pb+Pb collisions at LHC according to the MC Glauber model with $\sigma_{NN}=$ 64 mb.}
\label{table1.2}     
\end{center}
\end{table} 

The initial density profile is assumed to be proportional to the number of wounded nucleons
(WN), where the entropy density $s$ is distributed in the $(x,y)$ plane around the
wounded nucleons using a 2D Gaussian,
\begin{equation}
  s(x,y) = \frac{K}{2 \pi \sigma^2} \sum_{i=1}^{\ N_{\rm WN}} \exp \Big( -\frac{(x-x_i)^2+(y-y_i)^2}{2 \sigma^2} \Big).
 \label{eq:eps}
\end{equation}
We refer to this as the sWN profile. The parameter $K$ is a fixed overall normalization constant and
$\sigma$ is a free parameter determining the size of the fluctuations. The effective
interaction radius of the colliding nucleons, $\sqrt{\sigma_{NN}/\pi}/2 \ \sim$ 0.6~fm
(for RHIC), sets a natural order of magnitude for the size parameter. The default value
of $\sigma$ is taken as 0.4 fm from Ref.~\cite{hannu, chre}, however, we vary the value of $\sigma$ from 0.4 to 0.8 fm to check the sensitivity of the results to the fluctuation size.

Both intra-event fluctuations (i.e., 'hotspots' and 'holes') and the inter-event
fluctuations in the total entropy between different events are thus generated by the MCG
modeling of the initial state. We call events with larger than average entropy
"hot events" and those with smaller than average entropy "cold events".

The constant $K$ is taken as $102.1 \ \rm{fm^{-1}}$ for Au+Au collisions at
$\sqrt{s_{NN}} = 200 $ GeV. First the initial formation time ($\tau_0$)
of the plasma is fixed at $\tau_0 = 0.17$ fm/$c$ (as in Refs.~\cite{hannu, chre}),
motivated by the EKRT minijet saturation model~\cite{Eskola:1999fc} for
all the events. However, in the later part of this study we vary the value of $\tau_0$
 with collision centrality to check the effect of these variations and fluctuations in
the IC for thermal photon radiation. 
The freeze-out temperature is 160 MeV,
which reproduces the measured $p_T$ spectrum of pions~\cite{Adler:2003cb} well for the
sWN profile for a fixed $\tau_0$.

For the 2.76 ATeV Pb+Pb collisions at LHC we use the initial formation time $\tau_0 = 0.14$ fm/$c$ taken from EKRT model~\cite{risto}. 
The overall normalization constant in this case is  $K= 250 \ \rm{fm^{-1}}$ which reproduces the measured multiplicity of about 1600 \cite{alice}. The freeze-out temperature is kept same at RHIC and LHC.

\subsection{Thermal photon emission}
The quark-gluon Compton scattering and quark-anti-quark annihilation are the leading order processes for thermal photon production in the partonic phase. A significant contribution also comes from the resummed next to leading order bremsstrahlung processes~\cite{amy}. The rate of thermal photon production in the QGP depends on the momentum distribution of the partons which are governed by the thermodynamical conditions of the matter. 

The $\pi$ and $\rho$ mesons contribute dominantly to the photon production from a hot hadronic gas having temperature of the order of pion mass. This is due to the low mass of pions and the large spin iso-spin degeneracy of $\rho$ mesons, making them the most easily accessible particles in the medium~\cite{kls}. The leading photon producing channels involving  $\pi$ and $\rho$ mesons are $\pi \pi \ \rightarrow \ \rho \gamma$, $\pi \rho \ \rightarrow \ \pi \gamma$, and $\rho \ \rightarrow \ \pi \pi \gamma$.  

We use the rates $R=EdN/d^3p d^4x$ of~\cite{amy} for the plasma and those of~\cite{trg} for the hadronic matter which at present can be considered as the state of the art. The transition from the plasma rates to the hadronic rates is assumed to happen at a temperature of 170 MeV in this study.  The total thermal emission from the quark and hadronic matter phases is obtained integrating the rates over the space-time evolution of the medium, 
\begin{equation}
\label{eq1}
  E\, dN/d^3p = \int d^4x\, R\Big(E^*(x),T(x)\Big),
\end{equation}
where $E^*(x)=p^{\mu}u_{\mu}(x)$. The 4-momentum of the photon is 
$p^\mu{\,=\,}(p_T \cosh Y, p_T\cos\phi,p_T\sin\phi,p_T \sinh Y)$, and the 4-velocity of the flow field is $u^\mu = \gamma_T\bigl(\cosh \eta,v_x,v_y,$ $\sinh\eta\bigr)$ with $\gamma_T = (1{-}v_T^2)^{-1/2}$, $v_T^2{\,=\,}v_x^2{+}v_y^2$. The volume element is $d^4x{\,=\,}\tau\, d\tau \, dx \, dy\, d\eta$, where $\tau{\,=\,}(t^2{-}z^2)^{1/2}$ is the longitudinal proper time and $\eta{\,=\,}\tanh^{-1}(z/t)$ is the space-time rapidity. The photon
momentum is parametrized by its rapidity $Y$, transverse momentum $p_T$, and azimuthal emission angle $\phi$. 

\subsection{Averaging procedure}

The smooth initial density distribution from the MC Glauber calculation is obtained by
taking an average of 1000 fluctuating initial states~\cite{chre}. The  sWN
profile must be converted to energy density distribution using the EoS, since the input to
hydrodynamics is the energy density profile. When averaging over events in the initial
state, we apply the EoS in each event and in the end the average is taken over energy densities.
  Although photons are emitted during the entire space-time evolution,  the yield for $p_T > 1$ GeV/$c$ is dominated by QGP radiation. 
Thus averaging over 1000 events is enough to remove  the fluctuations from  the interior region of the fireball. 

The results from the fluctuating IC are obtained by averaging the photon spectra over a sufficiently large number of random events where the number needed increases for peripheral collisions. For example, we choose 40 events for 0--20\% centrality bin and 80 events for 40--60\% centrality bin. In order to fix the necessary number for all the centrality bins we check that addition of another very hot or very cold  event does not change the results significantly.

\begin{figure}
\centerline{\includegraphics*[width=8.0 cm]{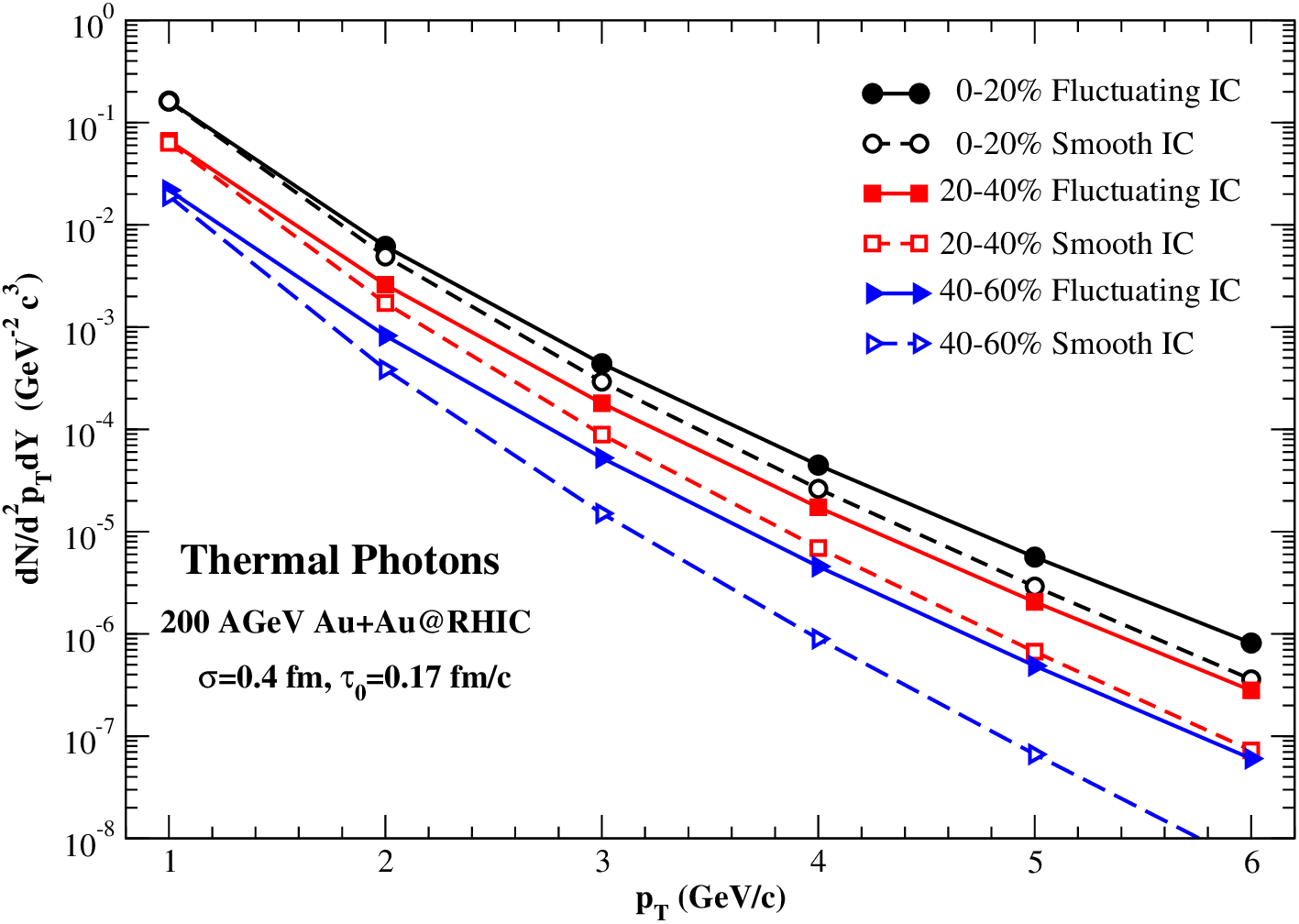}}
\caption{(Color online) Centrality dependence of thermal photon $p_T$ spectra for 200 AGeV Au+Au collisions at RHIC from fluctuating and smooth IC.}
\label{fig1}
\end{figure}
%
%

\section{Results from Fluctuating and smooth IC}
\subsection {Centrality dependence at RHIC}
First we study the centrality dependent results from 200 AGeV Au+Au collisions at RHIC considering $\tau_0=0.17$ fm/$c$.
Figure~\ref{fig1} shows the comparison between the smooth and the fluctuating IC for  thermal photon $p_T$ spectra  at different collision centralities  with $\sigma=0.4$ fm. The solid lines with closed symbols in the figure show the results from the  fluctuating IC and the dashed lines with open symbols are the results from a smooth initial state averaged IC (See Table~\ref{table1.1} for the $N_{\rm {part}}$ ranges and the corresponding values of average impact parameter in the different centrality bins).  

We see from Fig.~\ref{fig1} that the enhancement in the production from the fluctuating IC grows  for peripheral collisions  compared to the results from the smooth IC. The exponential slope of the spectra from the fluctuating IC is found to be about 10\%, 12\% and 16\% flatter than what is obtained from the smooth IC for 0 -- 20 \%, 20 -- 40 \% and 40 -- 60\%  centrality bins respectively, in the region $2 < p_T < 4$ GeV/$c$. Moving from central to peripheral collisions, the total number of 'hotspots' decreases in the initial density distribution as the number of participants is less for peripheral collisions. However, the presence of  even a few 'hotspots' in initial state makes significant enhancement in the production for peripheral collisions as the photon yield is very small there compared to the central collisions.  Thus the relative importance of initial state fluctuations increases towards peripheral collisions.

We have shown in the earlier work~\cite{chre} that production of thermal photons shows a very strong dependence on the  initial state density fluctuations. The 'hotspots' in the fluctuating IC enhance the production significantly in the range $p_T >$ 1 GeV/$c$ compared to the smooth initial state averaged IC for 0--20\% centrality bin at RHIC.  This is due to the exponential  temperature dependence of photon emission; the high $p_T$ photons are mostly emitted from the very early stage of the system expansion when hydrodynamic flow is still weak.  The low $p_T$ photons are mainly from the relatively cold later stage of the system expansion and are not effected significantly by the presence of 'hotspots' in the IC, as the strong pressure gradients remove the 'hotspots' rapidly.

\begin{figure}
\centerline{\includegraphics*[width=8.0 cm]{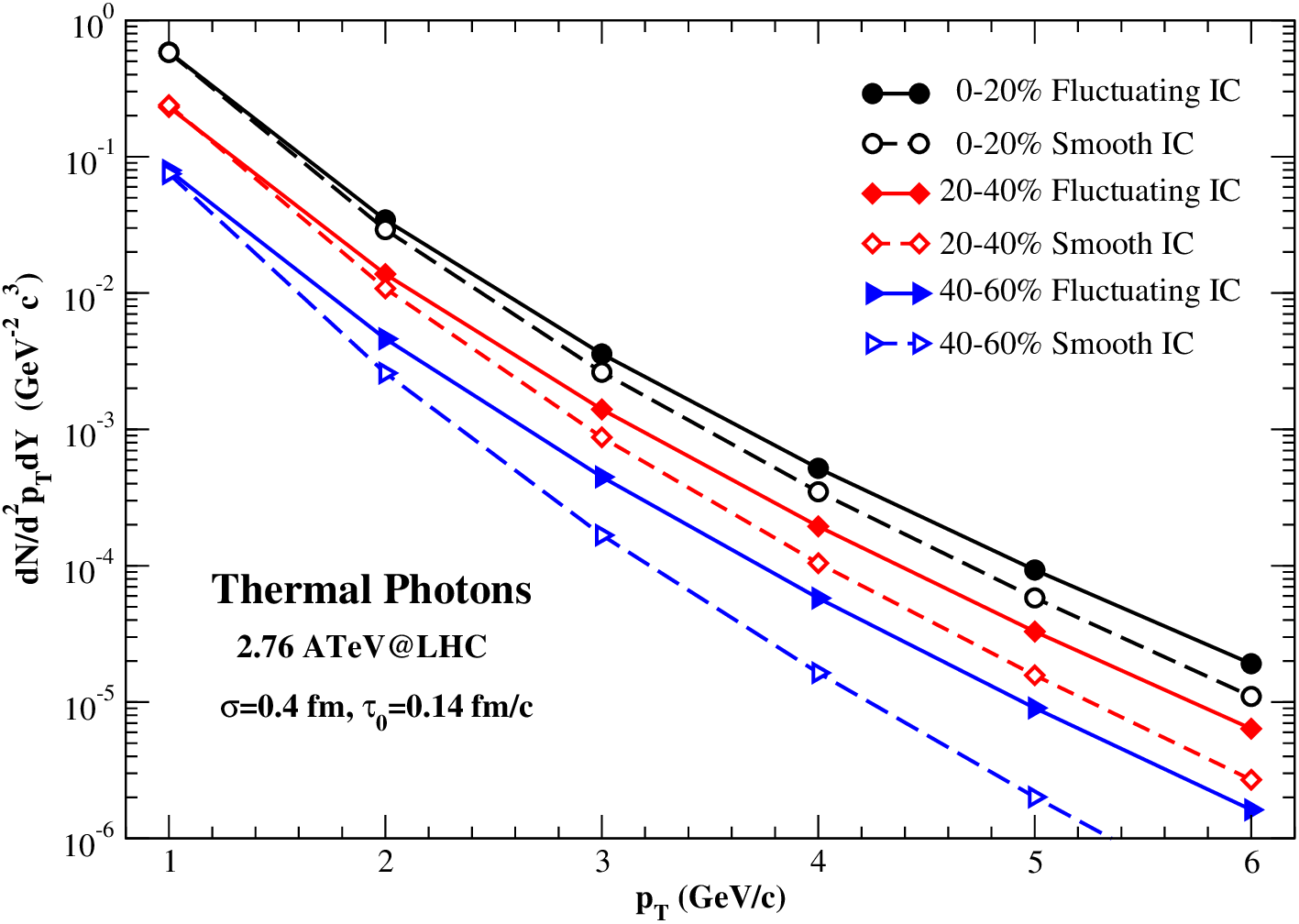}}
\caption{(Color online) Thermal photon $p_T$ spectra from smooth and fluctuating IC for 2.76 ATeV Pb+Pb collisions at LHC at different collision centralities  and at size parameter $\sigma=$ 0.4 fm.}
\label{fig2}
\end{figure}

We have checked that lowering the value of the freeze-out temperature from 160 to 120 MeV enhances the production significantly in the hadronic phase.  This additional contribution to the total spectrum, however, is not substantial for $p_T >$ 2  GeV/$c$ for central collisions. The relative contribution from the hadronic phase compared to the QGP phase increases towards peripheral collisions.  Thus, the lowering of the freeze-out temperature increases the production from hadronic phase in the low $p_T$ region more for peripheral collisions. However, fluctuations in the IC enhance the production even more for peripheral collisions and a  smaller value of freeze-out temperature  does not change results significantly in the thermal $p_T$ range.   It is to be noted that a bag model EoS produces more hadronic photons near the transition region compared to a lattice based EOS, however the contribution from the hadronic phase is not significant for $p_T > 2$ GeV/$c$ even for a bag model EOS~\cite{hre}.  In addition, we found  that these results are quite sensitive  to the initial profile details. The sWN profile leads to more pronounced maxima and hotter initial state than the energy initialization (eWN) profile~\cite{chre}. Thus also the thermal photon $p_T$ spectra are harder with the sWN profile.

\begin{figure}
\centerline{\includegraphics*[width=8.0 cm]{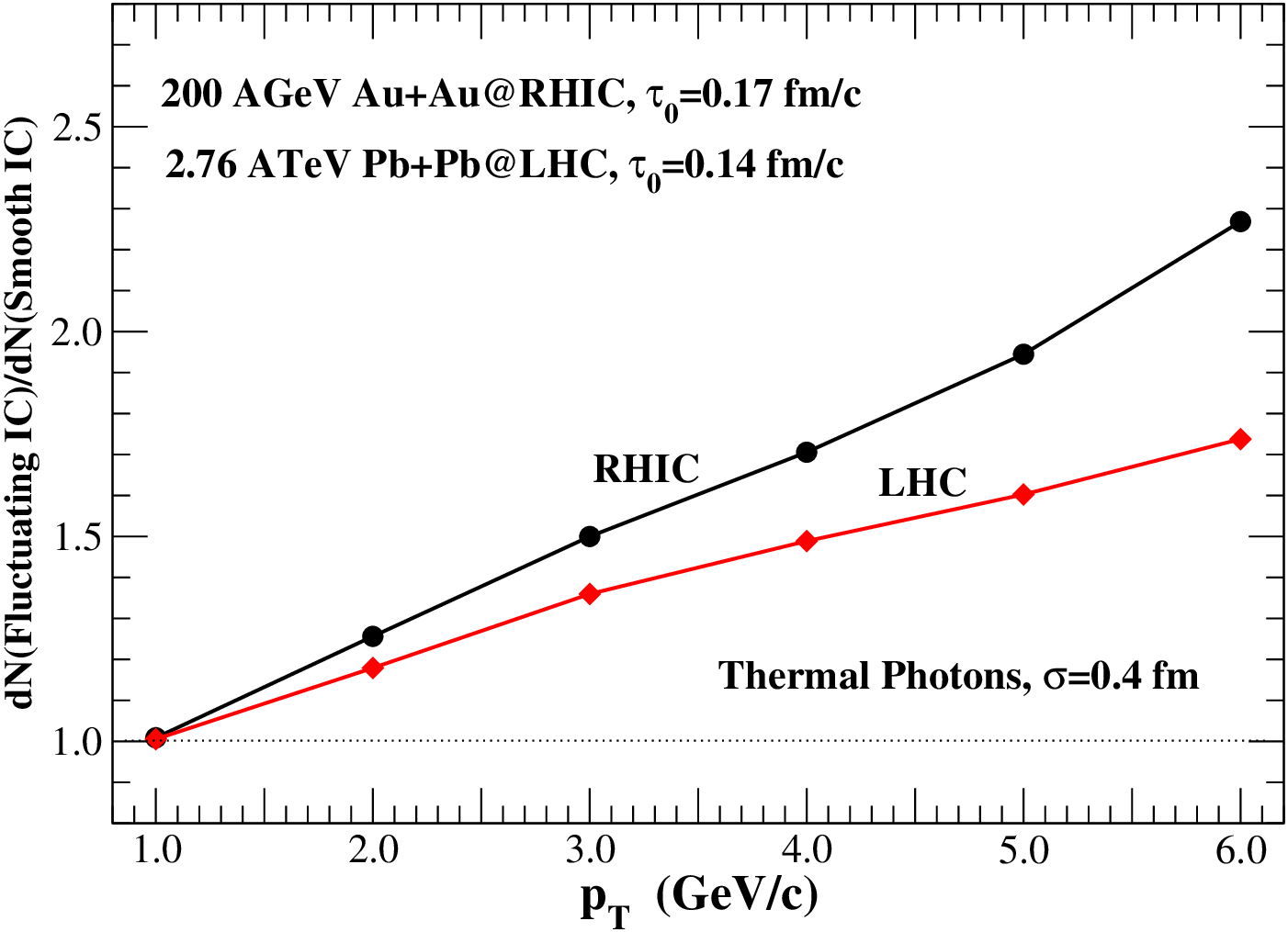}}
\caption{(Color online) Ratio of the thermal photon spectra from the smooth and the fluctuating IC for 0--20\% centrality bins at RHIC and LHC energies and for size parameter $\sigma=$ 0.4 fm.}
\label{fig3}
\end{figure}

\subsection{Results at LHC and comparison with RHIC results}
From the RHIC results at different collision centralities we saw that the effect of the initial state fluctuations is stronger when the system is smaller in size and the initial energy is less. Thus, one can expect that for Pb+Pb collisions at the LHC, where the initial energy density as well as the temperature is significantly higher than at RHIC, the effect of fluctuations will be less pronounced. 

Figure~\ref{fig2} shows the thermal photon spectra for different collision centralities at the LHC (See Table~\ref{table1.2} for the $N_{\rm part}$ ranges).  Results from the smooth IC are shown by the dashed lines with open symbols and those from the fluctuating IC  by the solid lines with closed symbols. The production of thermal photons is found to  increase by about a factor of 4 at $p_T$ =1 GeV/$c$ to a factor of more than 10 at $p_T$=4 GeV/$c$ from RHIC to LHC for the 0--20\% centrality bin. 
 The difference between the fluctuating and smooth initial conditions shows a
similar centrality dependence at LHC and at RHIC.

The ratio of thermal photon yield from fluctuating and smooth IC for 0--20\% centrality bin at the LHC is shown in the Fig.~\ref{fig3} where the result is compared with the RHIC result from Au+Au collisions at the same centrality bin. For both the energies the value of the size parameter $\sigma$ is taken as 0.4 fm.  As seen in Figs. \ref{fig1} and \ref{fig2}, fluctuations in the initial state enhance the production more for peripheral collisions than for central collisions. However, the overall magnitude of the enhancement is less at the LHC than at RHIC in the same $p_T$ range.

The results at LHC along with the centrality dependence results at RHIC confirm the earlier expectation that one might have to look at lower beam energies and smaller size systems for more sensitive measurements of the initial state fluctuations in heavy ion collisions.

\begin{figure}
\centerline{\includegraphics*[width=8.0 cm]{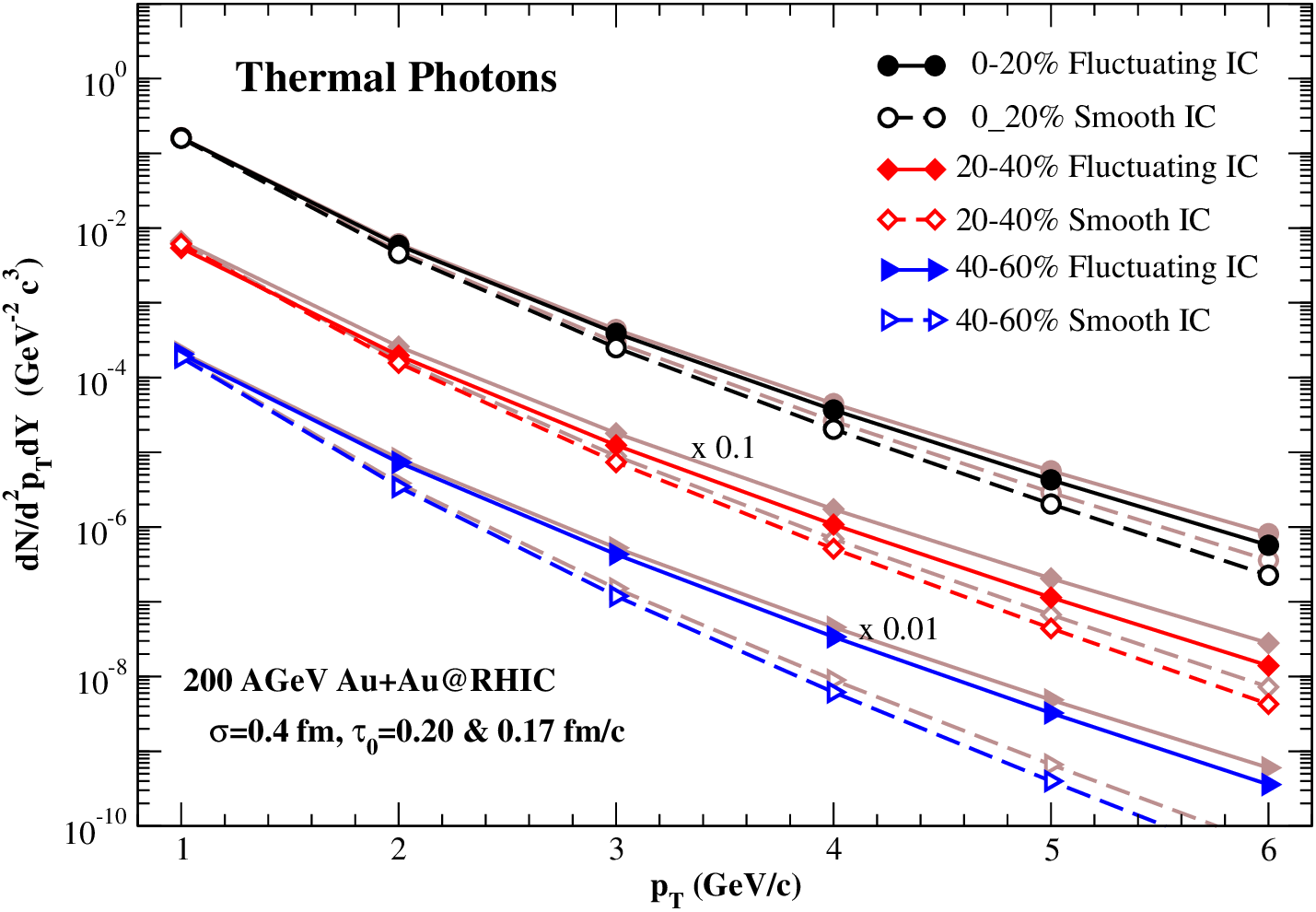}}
\caption{(Color online) Thermal photon $p_T$ spectra at different centrality bins for $\tau_0$ values of 0.17 fm/$c$ (gray lines) and 0.2 fm/$c$ and at a fixed value of size parameter $\sigma= 0.4$ fm. Results from fluctuating and smooth IC are shown by solid lines with closed symbols and dashed lines with open symbols respectively.}
\label{fig4}
\end{figure}

\subsection{Formation time dependence of the results at RHIC}
The initial formation time plays a very crucial role in the production of thermal photons from the plasma phase where a smaller  $\tau_0$ leads to a larger initial temperature and enhanced production of high $p_T$ photons~\cite{cs}. Thus, a significant enhancement in the photon production can also be obtained by reducing the value of the initial formation time for the smooth IC. 
As the value of the initial formation time is not known precisely, it is important to know whether the enhancement of the photon spectra due to fluctuations in the IC can be distinguished from  the enhancement due to a smaller value of the  formation time. A systematic study of thermal photon spectra at different centrality bins and at different values of $\tau_0$ is therefore necessary. 

Figure~\ref{fig4} shows the thermal photon spectra from the fluctuating and smooth IC at RHIC for two different values of $\tau_0$; 0.17 and 0.20 fm/$c$, where the value of $\tau_0$ is changed by keeping the total entropy of the system fixed.  We see that even a  small change  in the value of the formation time from 0.17 to 0.20 fm/$c$ affects the results from both the smooth and the fluctuating IC visibly. In the fluctuating case, the effect is stronger for the central collisions than for peripheral collisions as the relative enhancement in the production due to fluctuations is less for central collisions.  

The value of $\tau_0$ differs significantly in different  hydrodynamic models~\cite{hydro, uli} and mostly it varies from 0.17 to 0.60 fm/$c$ for  Au+Au collisions at maximum RHIC energy. We consider here 0.17 and 0.60 fm/$c$ as the lower and upper limits of  $\tau_0$ at RHIC and compare the photon spectra computed with these values. 

 When the initial time $\tau_0$ or the fluctuation  size parameter $\sigma$ is changed, also the
freeze-out temperature should be tuned in order to fit the hadron spectra. Here we
however keep the freeze-out temperature fixed for simplicity since the photon emission
from later stages of hadron gas phase is so small that total yield  in the $p_T$ range considered here would not change significantly. Note that this might not be true for other variables such as photon 
elliptic flow \cite{hre}.

\begin{figure}
\centerline{\includegraphics*[width=8.0 cm]{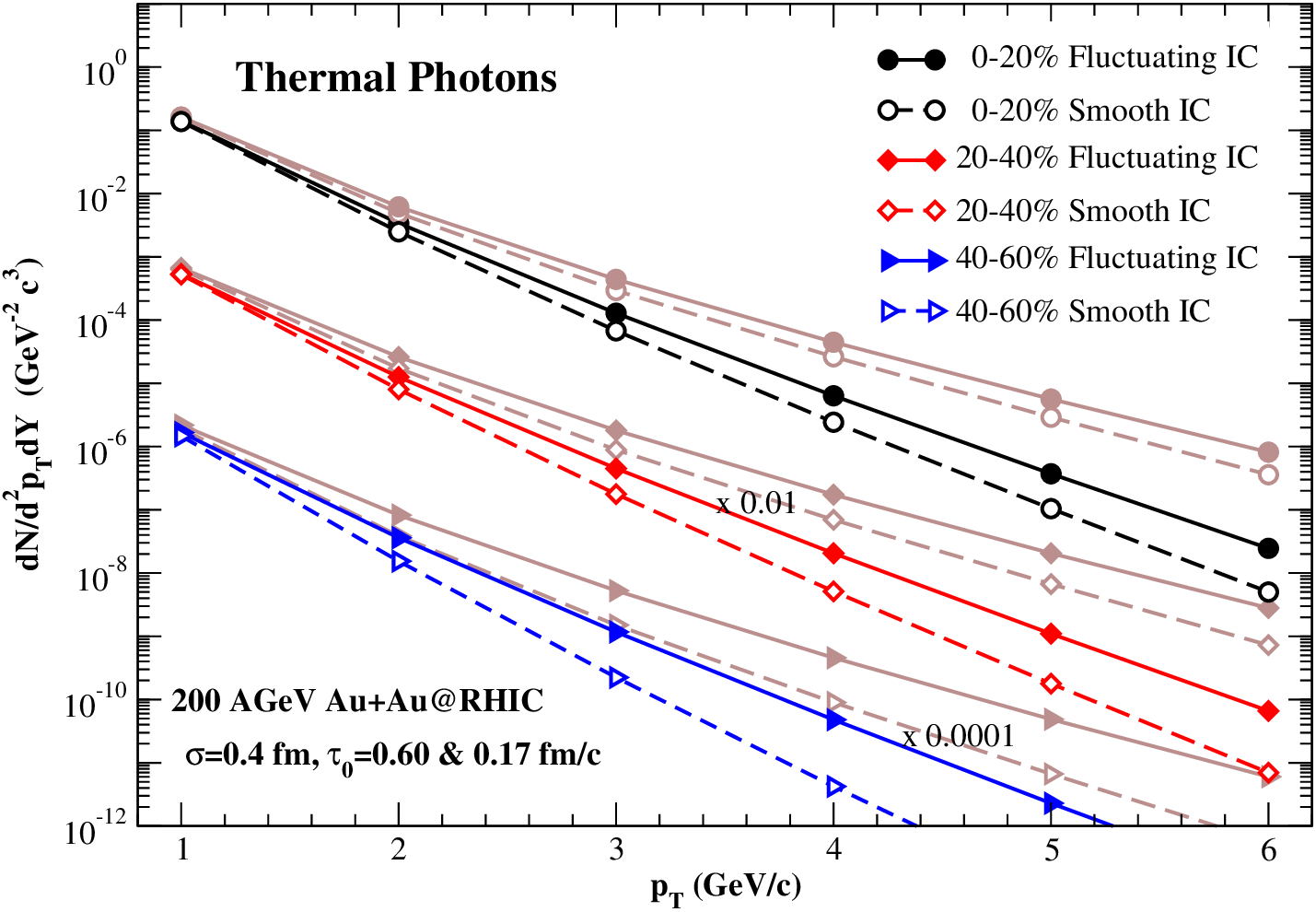}}
\centerline{\includegraphics*[width=8.0 cm]{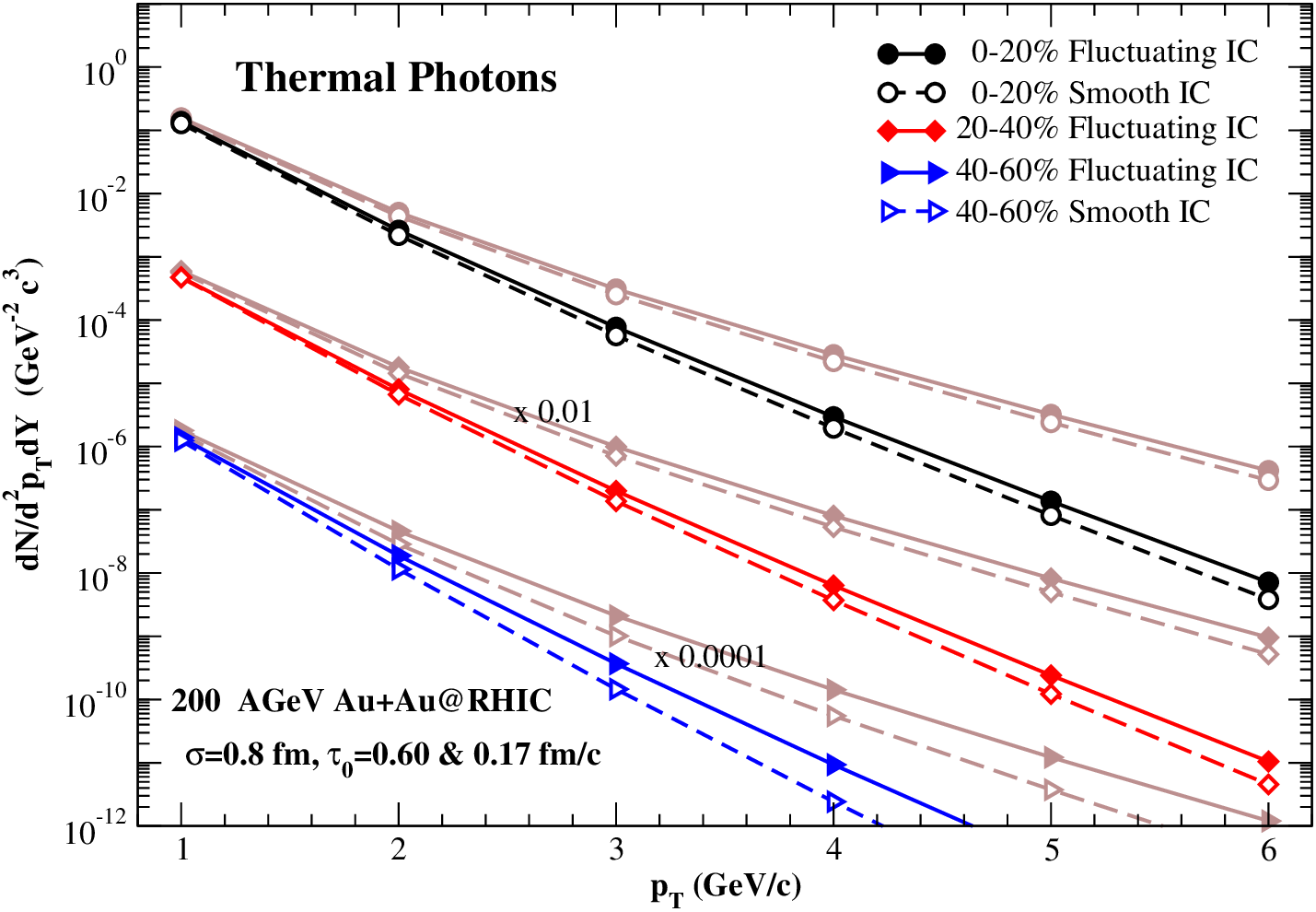}}
\caption{(Color online)Thermal photon $p_T$ spectra at different centrality bins for $\tau_0$ values of 0.17 fm/$c$ (gray lines) and 0.6 fm/$c$ (colored lines) and for size parameter $\sigma= 0.4$ (upper panel)  and $0.8$ (lower panel) fm. Results from fluctuating and smooth IC are shown by solid lines with closed symbols and dashed lines with open symbols respectively.}
\label{fig3.2}
\end{figure}

The size parameter $\sigma$ is a free parameter in our calculation and it is thus interesting to the see the effect of a simultaneous  change in the formation time and in the value of $\sigma$ on the production of thermal photons from the fluctuating IC. The upper panel of Fig.~\ref{fig3.2} shows the photon $p_T$ spectra from the fluctuating and smooth IC at $\tau_0$ values of 0.17 (gray lines) and 0.6 fm/$c$ (colored lines) for different collision centralities and  at a fixed value of the size parameter $\sigma=$ 0.4 fm. The lower panel of the same figure shows the results for the same parameters but with $\sigma=$ 0.8 fm. The yield falls sharply in the range $2 < p_T <4$ GeV/$c$ for all collision centralities as well as for both the $\sigma$ values when $\tau_0$ is increased from 0.17 to 0.60 fm/$c$. For a particular centrality bin, the difference between the smooth and the fluctuating IC is increased for larger values of $\tau_0$. This is again due to the fact that a larger value of $\tau_0$ leads to a relatively smaller value of initial temperature which is similar to a peripheral collision with smaller value of $\tau_0$.

These results show that it is difficult to conclude anything about the fluctuations in the IC and the size parameter $\sigma$ as long as $\tau_0$ is unknown.  

\section{Results with centrality dependent $\tau_0$}

So far we have considered different collision centralities by keeping the initial time $\tau_0$ fixed. 
However, in peripheral collisions the system is more dilute and the average energy scale of the produced quanta can be expected to be smaller than in central collisions, and thus both the system's formation and thermalization should take longer than in central collisions. Consequently, one should consider a $\tau_0$ which is systematically increasing from central to peripheral collisions \cite{Eskola:2000xq,Kolb:2001qz}. With the same logics, even within one collision, the $\tau_0$ should change according to the initial density fluctuations in the system. Clearly, such an advanced modeling of a fully local $\tau_0$ is not yet available. 

To get an order of magnitude estimate for the changes in the average $\tau_0$ from one centrality bin to another, we may, nevertheless perform an LHC case-study by using a simple non-local version of the EKRT model \cite{Eskola:1999fc}: First, as in \cite{Eskola:2001bf,hydro}, the average number of participants for each centrality bin is obtained from the optical Glauber model, and an effective mass number $A_{\rm eff}< A$ is determined so that the number of participants in a central $A_{\rm eff}$+$A_{\rm eff}$ collision matches with that of the centrality class in question. Then, an average $\tau_0$ for each centrality class is obtained by requiring that the production of minijets (gluons) of transverse momenta $p_T\ge p_0\gg\Lambda_{\rm QCD}$ in the mid-rapidity unit of the central $A_{\rm eff}$+$A_{\rm eff}$ collision fill the available effective transverse area,  
\begin{equation}
N_{A_{\rm eff} A_{\rm eff}}(b=0, p_0,\Delta y=1,\sqrt s) \cdot \pi/p_0^2 = \pi R_{A_{\rm eff}}^2, 
\end{equation}
where $R_{A_{\rm eff}}$ is the radius of the effective nucleus, $N_{A_{\rm eff} A_{\rm eff}}$ is the number of minijets computed in LO pQCD\footnote{Without any $K$-factors or an EKRT-based fit to the LHC multiplicity \cite{lhc_thorsten}, as we are only after a rough $\tau_0$ estimate here.}, and $\pi/p_0^2$ is the effective uncertainty area assigned to each minijet. The above condition then gives the saturation momentum as $p_0=p_{sat}$ and the average formation time of the system as $1/p_{sat}$. Assuming an instant thermalization -- which we believe is a reasonable approximation since the build-up of pressure and flow in the system starts very early -- one arrives at the $\tau_0$ values given in Table III\footnote{We thank Risto Paatelainen for computing these numbers for us from a work in progress \cite{risto}.}. For clarity, we note that the multiplicities at different centralities are still fixed here as explained in Secs. II. A. and III. C.; a more systematic study of multiplicities in the NLO EKRT framework will be presented elsewhere \cite{risto}. 

\begin{figure}
\centerline{\includegraphics*[width=8.0 cm]{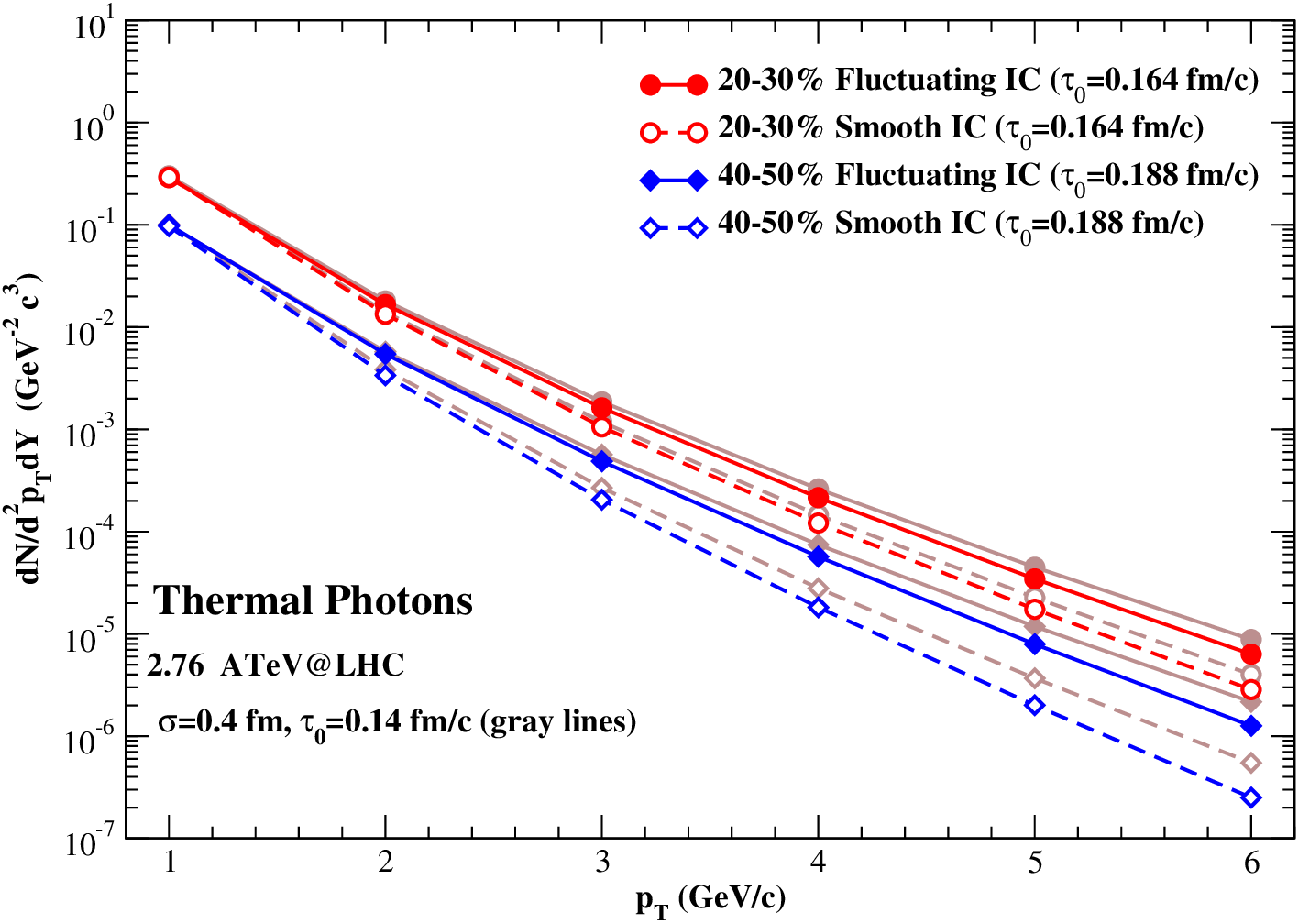}}
\caption{(Color online) Thermal photon spectra for peripheral collisions at the LHC from fluctuating and smooth IC considering centrality dependent  $\tau_0$ from the EKRT model.}
\label{lhc_tau}
\end{figure}

Table~\ref{table3} shows the obtained values of $\tau_0$, $A_{\rm eff}$ and $p_{sat}$ for three different centrality classes at LHC.  We notice that the change in $\tau_0$ from the 0-5 \% to the  40-50 \% centrality bin is only of the order of 30 \%, i.e. much less than the overall uncertainty (factor $\sim$ 3.5) in the $\tau_0$ considered in the previous section. 
Figure~\ref{lhc_tau} shows the corresponding thermal photon $p_T$ spectra at the LHC from the smooth and fluctuating IC for 
20-30 \%, and 40-50 \% centrality bins using the centrality-dependent $\tau_0$ along with the results obtained for a constant $\tau_0 = 0.14$~fm/c (as obtained for central collisions here). As we can see in the figure, for peripheral collisions and fixed $\tau_0$ the fluctuations in the IC enhance the production significantly (from the gray dashed lines to the gray solid lines). This enhancement is, however, partially compensated by a suppression caused by a larger formation time (from the gray solid lines to the colored solid lines). As a net result a visible enhancement relative to the case with smooth IC and fixed $\tau_0$ (from the dashed gray to the solid colored lines) remains.

From all the results at RHIC and LHC above, one can conclude that it is difficult to distinguish between the effects
of the IC fluctuations and formation time only by looking at the $p_T$ spectra of thermal photons. Instead, one needs to
look for some other experimentally measurable quantity where the ambiguities regarding the initial conditions can be reduced.

\begin{table} 
\begin{center}
    \begin{tabular}{ | c | c | c | c |}
    \hline
    Centrality bin (\%) & $p_0$ (GeV) & $A_{\rm{eff}}$  &  $\tau_0$ (fm) \\ \hline
    \hline 
    0 -- 5 & 1.39 & 193  &  0.140 \\ \hline
    20 -- 30 & 1.21 & 90  & 0.164 \\  \hline
    40 -- 50 & 1.05 & 40  & 0.188 \\ \hline
    \end{tabular}
\caption{value of $\tau_0$ for there different centrality bins for 2.76 ATeV Pb+Pb collisions at LHC  calculated from the EKRT model (see Ref.~\cite{Eskola:1999fc} for details).} 
\label{table3} 
\end{center}
\end{table}

%
%
\begin{table}
\begin{center}
    \begin{tabular}{ | c | c | c | c | c |}
    \hline
     Centrality \% & $N_{\rm{part}}$ range & $\langle b \rangle  $ (fm) & $\langle N_{\rm{part}} \rangle$ & $\langle N_{\rm{bin}} \rangle$\\ \hline
     \hline
      0 -- 10 & 394 -- 276 & 3.14 & 326.3 & 958.9 \\ \hline
     10 -- 20 & 275 -- 197 & 5.72 & 234.0 & 599.0 \\  \hline
     20 -- 30 & 196 -- 138  & 7.40 & 165.6 & 369.6 \\ \hline
     30 -- 40 & 137 -- 93   & 8.76 & 113.9 & 217.8 \\ \hline
     40 -- 50 & 92 -- 59 & 9.94 & 74.55 & 119.9\\ \hline
     50 -- 60 & 58 -- 34 & 11.0 & 45.20 & 59.74 \\  \hline

    \end{tabular}	
\caption{$N_{\rm{part}}$, average impact parameter, average number of participants and average number of binary collisions for different centrality bins and for  200 AGeV Au+Au collisions at RHIC with $\sigma_{NN}=$ 42 mb.}
\label{table1.3}     
\end{center}
\end{table}

\subsection{Ratio of central to peripheral yield at RHIC}
The ratio of direct photon $p_T$ spectra from central to peripheral collisions is  a useful measure since all overall normalization factors (which may not be well known) are cancelled. 
This ratio as a function of collision centrality also offers the possibility to better focus on the system size dependence of the IC fluctuation.

We define a quantity $R_{cp}^{\gamma}$, which is the ratio of central to peripheral yield of thermal photons normalized by the number of binary collisions
\begin{equation}
R_{cp}^{\gamma}|_i= \frac {dN/d^2p_TdY|_{0-10 \%}} {dN/d^2p_TdY|_{i-j \%}} \times \frac  {N_{bin}|_{i-j\%}} {N_{bin}|_{0-10 \%}},
\end{equation}
where the value of $i$ is changed from 10 to 70 in steps of 10 and $j$=$i$+10 in the following. Note that although not considered here, for pQCD direct photons such ratio is close to unity. 
We calculate $R_{cp}^{\gamma}$ as a function of collision centrality for different values of transverse momentum  as well as a function of the size parameter $\sigma$.

Note that we choose a different definition of $R_{cp}^{\gamma}$ rather than the conventional definition where the result from the most peripheral collision is kept as a fixed denominator and  the numerator is changed for different centrality bins. The hydrodynamical framework is more realistic for central collisions than for peripheral collisions 
and as a result we keep the numerator fixed and change the denominator accordingly.

\begin{figure}
\centerline{\includegraphics*[width=8.0 cm]{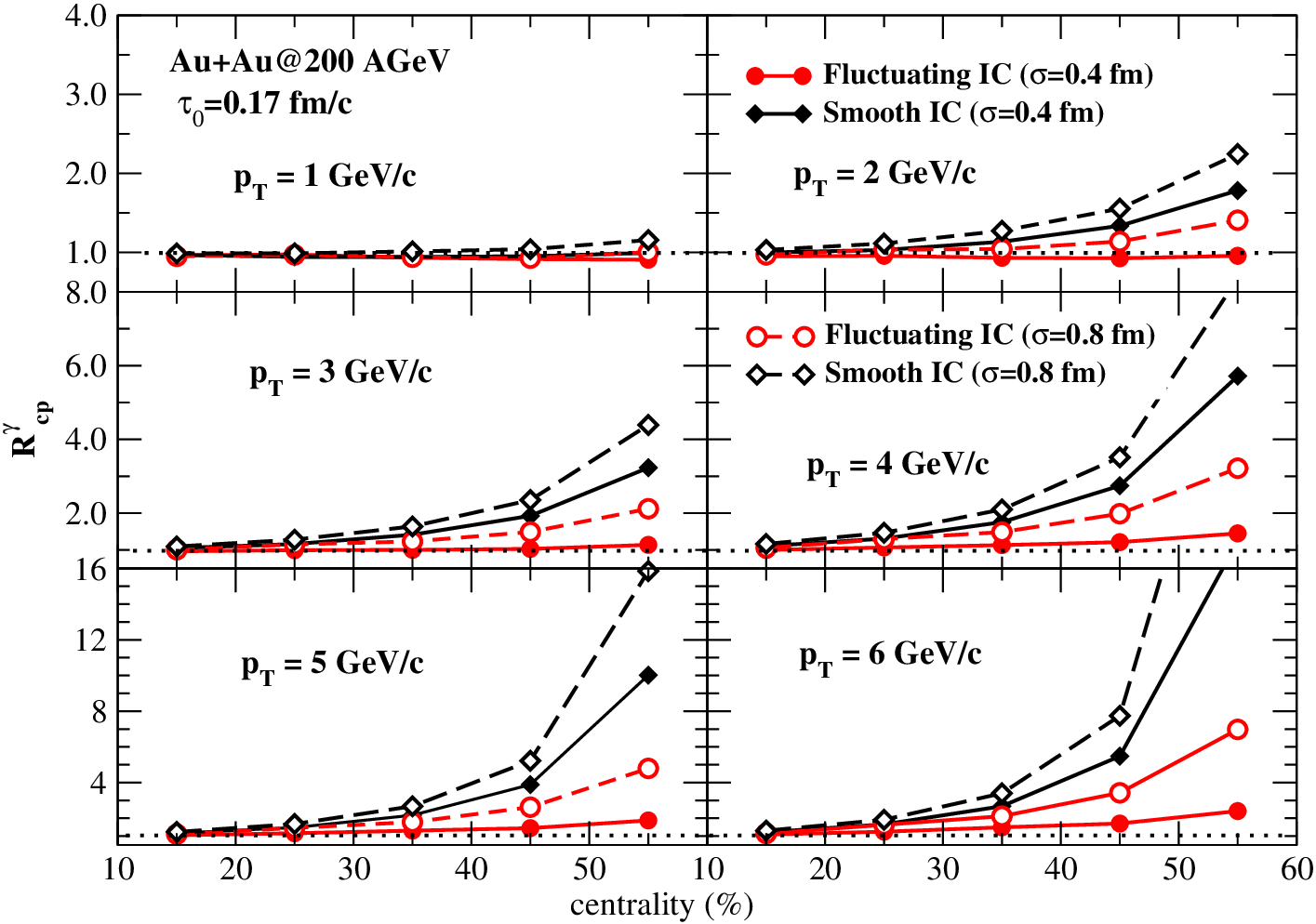}}
\caption{(Color online) $R_{cp}^{\gamma}$ as a function of collision centrality for thermal photons at different $p_T$ values for 200 AGeV Au+Au collisions at RHIC. Two different values of $\sigma$,  0.4 and 0.8 fm are used. The $N_{\rm part}$ ranges for the different centrality bins are taken from Table~\ref{table1.3}.}
\label{rcp1}
\end{figure}
Figure~\ref{rcp1} shows the variation of $R_{cp}^{\gamma}$ as a function of collision centrality for different values of transverse momentum and for $\sigma$ values of 0.4 and 0.8 fm.
The results both from the smooth and the fluctuating IC approach  1 towards central collision  by definition and for a fixed $p_T$ a clear difference between the results from the fluctuating and smooth IC can be observed for peripheral collisions. 

At $p_T=$ 1 GeV/$c$,  differences between the results from the fluctuating and the smooth IC are small for central collisions and they increase slightly for peripheral collisions. In the thermal region $2 \le p_T \le 4$ GeV/$c$, $R_{cp}^{\gamma}$ from the fluctuating IC does not change significantly with collision centrality. In contrast, for smooth IC where only the $\tau_0$ dependence is relevant, the value of $R_{cp}^{\gamma}$ rises rapidly with collision centrality for both values of $\sigma$. Fluctuations in the IC enhance the production more for peripheral collisions compared to the smooth profile case.  As a result the central to peripheral ratio for the smooth IC is always larger than for the fluctuating IC and the difference increases towards the larger values of transverse momentum. 

For larger values of $\sigma$, the effect of fluctuations becomes less pronounced and as a result $R_{cp}^{\gamma}$ for peripheral collisions increases with the centrality at large $p_T$. Results from the smooth IC rise faster compared to the fluctuating IC at $\sigma=$ 0.8 fm indicating that some fluctuation effects are still present even with this $\sigma$ value.
\begin{figure}
\centerline{\includegraphics*[width=8.0 cm]{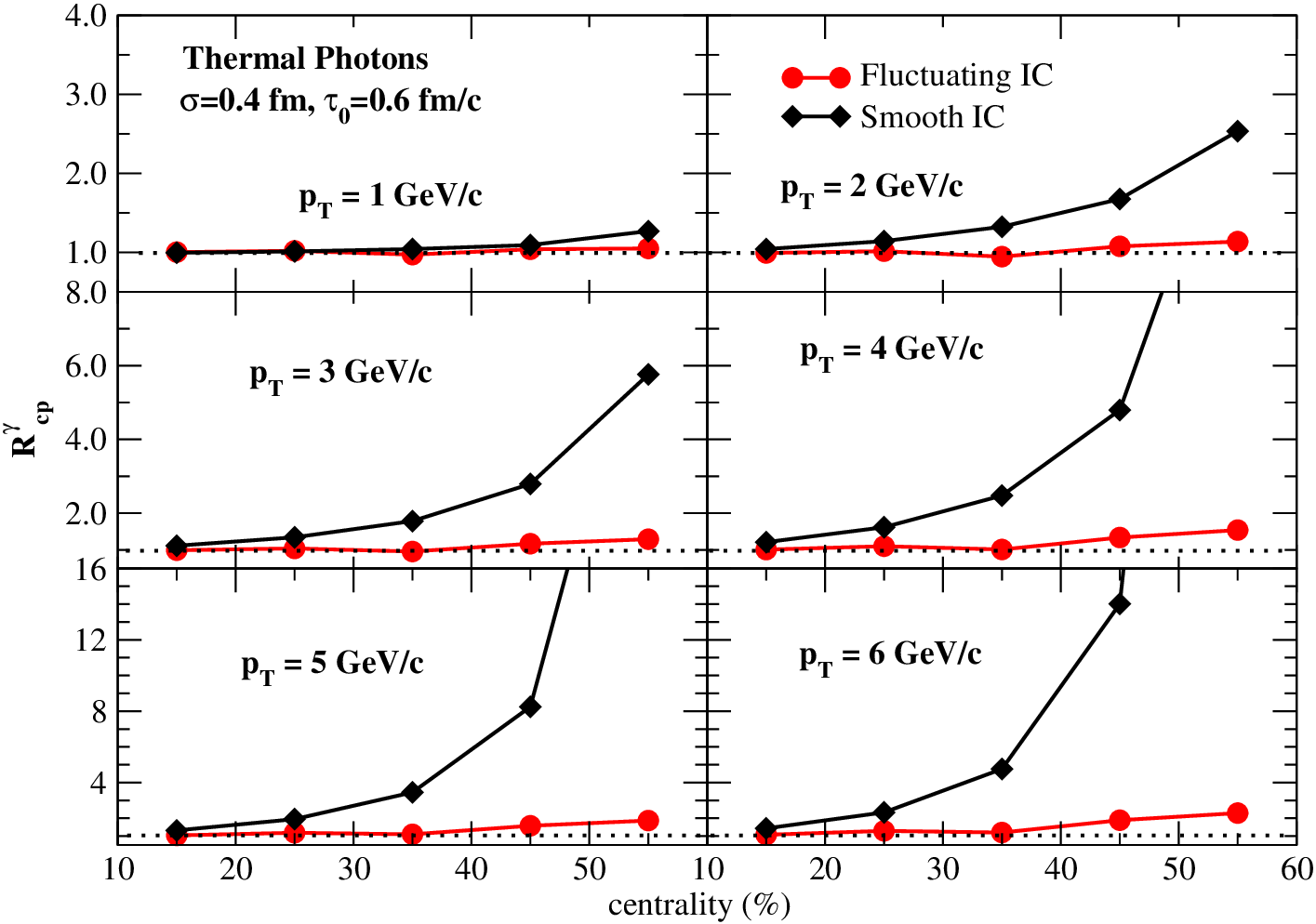}}
\caption{(Color online) $R_{cp}^{\gamma}$ as a function of collision centrality for thermal photons at different $p_T$ values for 200 AGeV Au+Au collisions at RHIC. Size parameter $\sigma$ is kept fixed at 0.4 fm.}
\label{fig5.3}
\end{figure}

\begin{figure}
\centerline{\includegraphics*[width=8.0 cm]{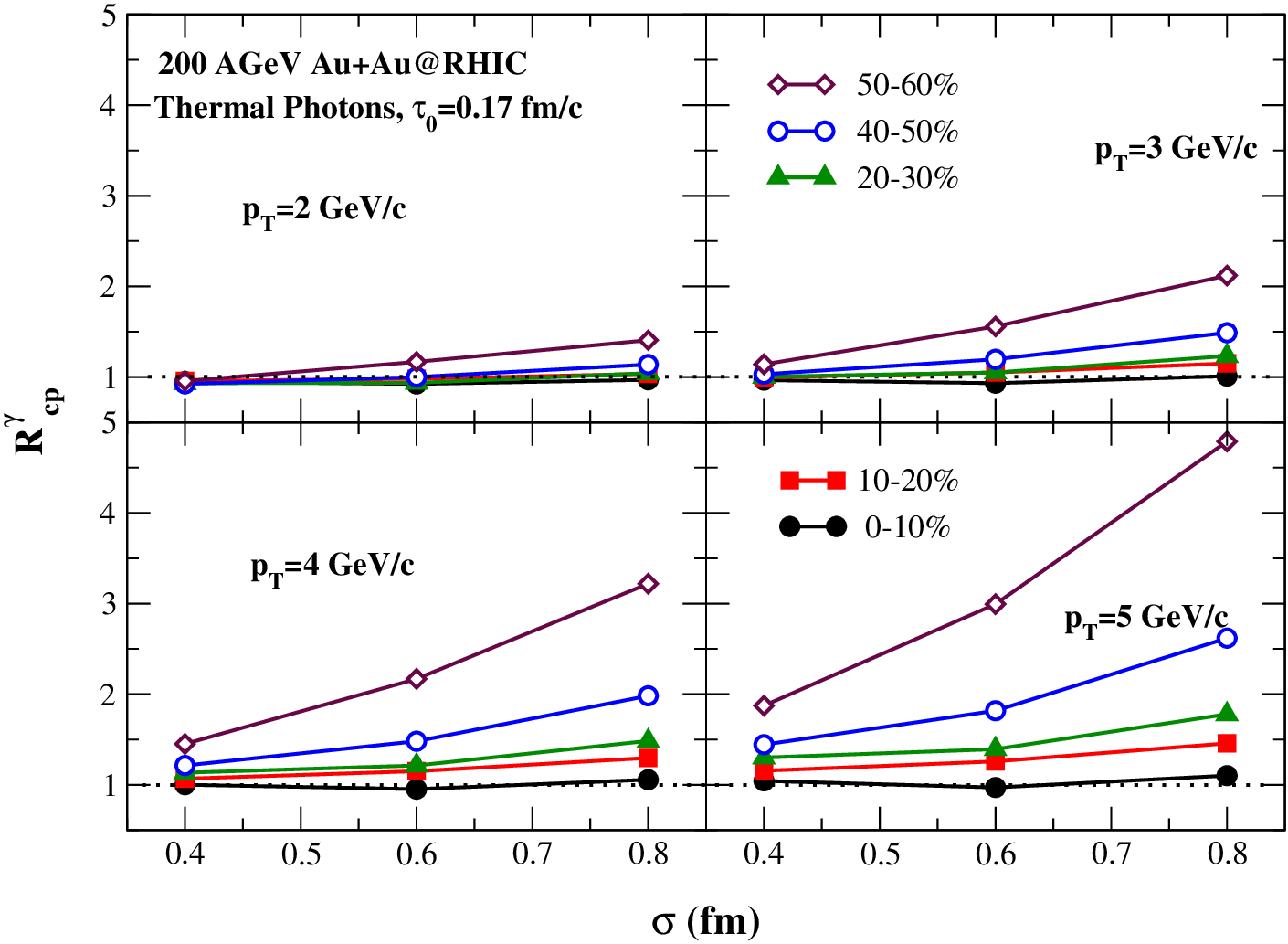}}
\caption{(Color online) $R_{cp}^{\gamma}$  at $p_T$ values of 2, 3, 4, and 5 GeV/$c$ for 200  AGeV Au+Au collisions at RHIC as a function of size parameter $\sigma$ from fluctuating IC.}
\label{fig10}
\end{figure}
In Fig.~\ref{fig5.3}, $R_{cp}^{\gamma}$ for thermal photons as a function of collision centrality is shown for  $\tau_0$ = 0.6 fm/$c$ and at different values of $p_T$.  
For the smooth IC, the value of $R_{cp}^{\gamma}$ increases rapidly from central towards peripheral collisions and the increase is more for larger values of transverse momentum, whereas for the fluctuating IC the variation of $R_{cp}^{\gamma}$ with collision centrality is not very rapid, and only for a very large $p_T$ a significant increase in the value of $R_{cp}^{\gamma}$ can be observed. Thus, comparing Figs.~\ref{rcp1} and~\ref{fig5.3}, it can be seen that the difference between the $R_{cp}^{\gamma}$ values from the smooth and fluctuating profiles is more pronounced for larger $\tau_0$. These results show that  $R_{cp}^{\gamma}$ is a potentially useful observable in the study of IC fluctuations and the $\tau_0$ uncertainty. 

The $R_{cp}^{\gamma}$ results for four different values of transverse momentum are shown in Fig.~\ref{fig10} as a function of the size parameter $\sigma$ at different collision centralities. 
For $p_T$ =2 GeV/$c$ $R_{cp}^{\gamma}$ shows very little variation with  $\sigma$ for almost all the centralities as the 'hotspots' in the fluctuating IC affect the spectra marginally in this $p_T$ range. However for $p_T = 3$ GeV/$c$,  $R_{cp}^{\gamma}$ shows stronger dependence on $\sigma$ as fluctuations in the IC affect the spectra more significantly and the effect is stronger for  smaller values of $\sigma$. A much larger variation of $R_{cp}^{\gamma}$ with $\sigma$ can be seen for $p_T\ge 4$ GeV/$c$ and even for central and mid-central collisions. We have good reasons to believe that the direct photon spectra in the $p_T$ range 3--4 GeV/$c$  is dominated by thermal radiation~\cite{chre} and the photon spectra can be measured experimentally up to a very large collision centrality as shown in the direct photon data by PHENIX. Thus, an experimental measurement of the parameter $R_{cp}^{\gamma}$ at different $p_T$ values and at different  collision centralities can actually serve as a probe for the size parameter $\sigma$.

\begin{table}
\begin{center}
    \begin{tabular}{ | c | c | c | c | c |}
    \hline
     Centrality \% & $N_{\rm {part}}$ range & $\langle b \rangle  $ (fm) & $\langle N_{\rm{part}} \rangle$ & $\langle N_{\rm{bin}} \rangle$ \\ \hline
     \hline
      0 -- 10 & 416 -- 306 & 3.27 & 357.5 & 1506 \\ \hline
     10 -- 20 & 305 -- 221 & 6.00 & 261.0 & 929.0 \\  \hline
     20 -- 30 & 220 -- 155  & 7.78 & 186.1 & 562.3 \\ \hline
     30 -- 40 & 154 -- 104   & 9.26 & 127.7 & 319.1 \\ \hline
     40 -- 50 & 103 -- 66 & 10.5 & 83.36 & 168.8\\ \hline
     50 -- 60 & 65 -- 39 & 11.6 & 51.34 & 82.37 \\  \hline
    \end{tabular}	
\caption{$N_{\rm{part}}$, average impact parameter, average number of participants and average number of binary collisions for different centrality bins and for  2.76 ATeV Pb+Pb collisions at LHC with $\sigma_{NN}= $ 64 mb.}
\label{lhctab}     
\end{center}
\end{table}

\begin{figure}
\centerline{\includegraphics*[width=8.0 cm]{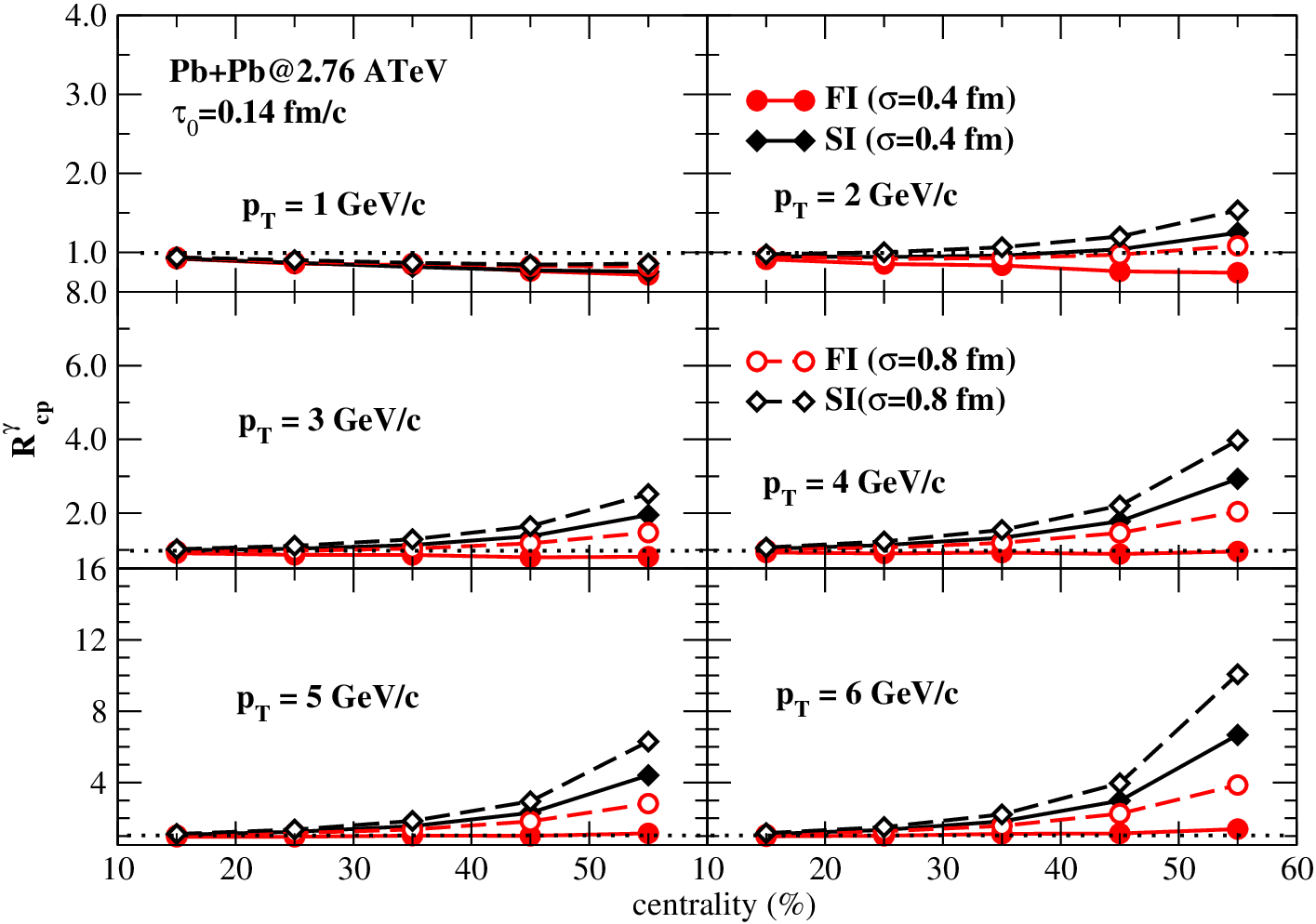}}
\caption{(Color online) $R_{cp}^{\gamma}$ as a function of collision centrality for thermal photons at different $p_T$ values for 2.76 ATeV Pb+Pb collisions at LHC. Two different values of $\sigma$,  0.4 and 0.8 fm are used. $N_{\rm part}$ ranges are taken from Table \ref{lhctab}.}
\label{fig11}
\end{figure}

Finally, in Fig.~\ref{fig11} we plot the central to peripheral ratio at LHC (see Table~\ref{lhctab} for the $N_{\rm{part}}$ and $N_{\rm{bin}}$ ranges for different centrality bins at LHC). We notice that the ratio $R_{cp}^{\gamma}$ is generally much flatter at higher collision energy. Also the difference between the smooth and fluctuating initial conditions is smaller at the LHC than at RHIC.
 
\section{Summary and conclusions}
In conclusion, we have shown that fluctuations in the IC lead to a significant enhancement in the production of thermal photons compared to a smooth initial state averaged IC in the ideal hydrodynamic calculation. For a particular beam energy the  enhancement is found to be  more pronounced for peripheral collisions than for central collisions.  The relative enhancement is found to be comparatively less at LHC than at RHIC for the same centrality bin.

The thermal photon $p_T$ results at RHIC and LHC are found to be quite sensitive to the value of the initial formation time. A larger value of formation time  lowers the production of thermal photons generically.  One might also expect that the value of  $\tau_0$  is not fixed and it may change with collision centralities. We have shown the $p_T$ spectra from the smooth and fluctuating IC considering different values of the initial formation time as well as results considering centrality dependent $\tau_0$ values from the  EKRT model. The effect of fluctuations in the IC and a larger value of formation time act in opposite directions   by  increasing and decreasing the production respectively. All these results suggest that the effect of the IC fluctuations and change in the  formation time  can not be distinguished easily only by studying the spectra of thermal photons.

We have shown that the central to peripheral ratio of $p_T$ spectra (normalized with the number of binary collisions) can be a useful measure to reduce the model-dependent uncertainties in the calculations.  Experimental determination of $R_{cp}^{\gamma}$  at different $p_T$  and at different collision centralities  would be a valuable tool for  estimating the size scale of the fluctuations. The difference between the $R_{cp}^{\gamma}$ results from the  smooth and the fluctuating IC has been found stronger for peripheral collisions and for larger values of $p_T$. We also found that at the LHC energy the $R_{cp}^{\gamma}$ is much flatter than at RHIC and the difference between fluctuating and smooth IC is smaller. 
 However, one should bare in mind that  hydrodynamics is not expected to work at very peripheral collisions ($ > 60$\%), and that at  large $p_T$ ($> 4 $ GeV/$c$)  the pQCD photons dominate the direct photon spectra.

\begin{acknowledgments} 
We thank R. Paatelainen for providing us with his EKRT code and  the $\tau_0$ values for peripheral collisions at LHC. We gratefully acknowledge the financial support by the Academy of Finland. TR and RC are supported by the Academy researcher program (project  130472) and KJE by a research grant (project 133005). In addition, HH was supported by the national Graduate School of Particle and Nuclear Physics and the Extreme Matter Institute (EMMI). We acknowledge CSC -- IT Center for Science in Espoo, Finland, for the allocation of computational resources.

\end{acknowledgments}


\begin{thebibliography}{99}
\bibitem{fl2} C.~Adler {\it et al.} [STAR Collaboration], Phys.\ Rev.
\ Lett. {\bf 87}, 182301 (2001); ibid {\bf 89}, 132301 (2002); ibid
{\bf 90}, 032301 (2003); S.~S.~Adler {\it et al.} [PHENIX Collaboration],
Phys.\ Rev. \ Lett. {\bf 91}, 182301 (2003).

\bibitem{fl1} P.~Huovinen, P.~Kolb, U.~Heinz, P.~V.~Ruuskanen, and
S.~Voloshin, \ Phys.\ Lett. \ B {\bf 503}, 58 (2001); D.~Teaney, J.~Lauret,
 and E.~Shuryak, nucl-th/0110037.

\bibitem{Niemi:2008ta}
H.~Niemi, K.~J. Eskola, and P.~V. Ruuskanen,
\newblock Phys. Rev. {\bf C79}, 024903 (2009).

\bibitem{hydro}  K.~J.~Eskola, H.~Honkanen, H.~Niemi, P.~V.~Ruuskanen and S.~S.~Rasanen,
Phys.\ Rev.\  C {\bf 72}, 044904 (2005);
  P.~Huovinen and P.~V.~Ruuskanen,
 Ann.\ Rev.\ Nucl.\ Part.\ Sci.\  {\bf 56}, 163 (2006);
   C.~Nonaka and S.~A.~Bass,
 Phys.\ Rev.\  C {\bf 75}, 014902 (2007).

\bibitem{hirano}
  T.~Hirano,
  Phys.\ Rev.\ C {\bf 65}, 011901 (2002);
  T.~Hirano and K.~Tsuda,
  Phys.\ Rev.\ C {\bf 66}, 054905 (2002).


\bibitem{uli} 
  P.~F.~Kolb and U.~W.~Heinz, Quark gluon plasma 3, ed. R. C. Hwa and X. N. Wang (Singapore: World Scientific) 634 (2003)
  [nucl-th/0305084].


\bibitem{hannu} H. Holopainen, H. Niemi, and K. J. Eskola, Phys. \ Rev. \ C {\bf 83}, 034901 (2011). 

\bibitem{hama}
Y.~Hama, T.~Kodama, and O.~Socolowski, Jr.,
\newblock Braz. J. Phys. {\bf 35}, 24 (2005).

\bibitem{andrade} R.~Andrade, F.~Grassi, Y.~Hama, T.~Kodama, and O.~Socolowski, Jr., Phys. Rev. Lett. {\bf 97}, 202302 (2006);
R.~P.~G. Andrade, F.~Grassi, Y.~Hama, T.~Kodama, and W.~L. Qian, Phys. Rev. Lett. {\bf 101}, 112301 (2008).

\bibitem{phot} 
  P.~V.~Ruuskanen,
  Nucl.\ Phys.\  A {\bf 544}, 169 (1992), and references therein.

\bibitem{dks_qm08} D. K. Srivastava, \ J. \ Phys. \ G {\bf 35}, 104026 (2008).

\bibitem{phenix1} A. Adare {\it et al.} [PHENIX Collaboration], Phys.  \ Rev. \ Lett. {\bf 104}, 132301 (2010).

\bibitem{amy} P.~Arnold, G.~D.~Moore, and L.~G.~Yaffe, JHEP {\bf 0112}, 009
(2001).

\bibitem{trg} S.~Turbide, R.~Rapp, and C.~Gale, \ Phys. \ Rev. \ C {\bf 69},
 014903 (2004).

\bibitem{chre}
  R.~Chatterjee, H.~Holopainen, T.~Renk, and K.~J.~Eskola,
  Phys.\ Rev.\  {\bf C83}, 054908 (2011); 
  R.~Chatterjee, H.~Holopainen, T.~Renk, K.~J.~Eskola,
  J. \ Phys. \ G {\bf 38}  124136 (2011).



\bibitem{Eskola:1999fc}
K.~J. Eskola, K.~Kajantie, P.~V. Ruuskanen, and K.~Tuominen,
\newblock Nucl. Phys. {\bf B570}, 379 (2000).
\bibitem{risto} R. Paatelainen, H. Holopainen, and  K. J. Eskola [in progress].
\bibitem{Boris}
J.~P. Boris and D.~L. Book,
\newblock J. Comput. Phys. {\bf A11}, 38 (1973).

\bibitem{Zalesak}
S.~T. Zalesak,
\newblock J. Comput. Phys. {\bf A31}, 335 (1979).

\bibitem{Laine:2006cp}
M.~Laine and Y.~Schroder,
\newblock Phys. Rev. {\bf D73}, 085009 (2006).

\bibitem{Adler:2003cb}
S.~S. Adler {\em et~al.},
\newblock Phys. Rev. {\bf C69}, 034909 (2004).
\bibitem{alice} K. Aamodt {\it et al.} [ALICE Collaboration], Phys. \ Rev. \ Lett. {\bf 105}, 252301 (2010).

\bibitem{kls} J.~I.~Kapusta, P.~Lichard and D.~Seibert,
  Phys.\ Rev.\  D {\bf 44}, 2774 (1991)
  [Erratum-ibid.\  D {\bf 47}, 4171 (1993)]
  [Phys.\ Rev.\  D {\bf 47}, 4171 (1993)].
 

\bibitem{hre}	
H. Holopainen, S. S. Rasanen, and K. J. Eskola,
 \ Phys. \ Rev. \ C {\bf84}, 064903 (2011); M.~Dion, C.~Gale, S.~Jeon, J.~-F.~Paquet, B.~Schenke and C.~Young,
  J.\ Phys.\  G {\bf 38}, 124138 (2011).




\bibitem{cs} R. Chatterjee and D. K. Srivastava, Phys. \ Rev. \ C {\bf 79}, 021901(R) (2009); R. Chatterjee and D. K. Srivastava, Nucl. \ Phys. {\bf A830}, 503c (2009). 




\bibitem{Eskola:2000xq}
  K.~J.~Eskola, K.~Kajantie and K.~Tuominen,
  Phys.\ Lett.\  B {\bf 497}, 39 (2001).
\bibitem{Kolb:2001qz}
  P.~F.~Kolb, U.~W.~Heinz, P.~Huovinen, K.~J.~Eskola and K.~Tuominen,
  Nucl.\ Phys.\  A {\bf 696}, 197 (2001).

\bibitem{Eskola:2001bf} 
  K.~J.~Eskola, P.~V.~Ruuskanen, S.~S.~Rasanen and K.~Tuominen,
  Nucl.\ Phys.\ A {\bf 696}, 715 (2001).

\bibitem{lhc_thorsten}
T. Renk, H. Holopainen, R. Paatelainen, and K. J. Eskola;
Phys. \ Rev.  \ C {\bf 84}, 014906 (2011).

\end{thebibliography}
\end{document}